# Phase stability and Nucleation effect upon Iron addition to SmCo$_5$ bulk magnets

B. Das[*], B. Balamurugan, and D. J. Sellmyer

*Nebraska Center for Materials and Nanoscience and Department of Physics and Astronomy*

*University of Nebraska, Lincoln, NE 68588*

[*] *bhaskar.das@huskers.unl.edu*

**Abstract:**

Change in structure and magnetic properties of $SmCo_{5-x}Fe_x$ permanent magnets for different values of x ranging from 0 to 2 have been studied. Structural investigation from X-ray diffraction (XRD) patterns confirms the hexagonal $CaCu_5$ –type structure of the $SmCo_{5-x}Fe_x$ ribbons for $0 \leq x < 2$. The decrease in angular position of the diffraction peaks points to the lattice expansion due to the substitution of Co atoms by larger Fe atoms. Mixture of phases occurs for $x \geq 2$ and has been confirmed by both XRD studies and magnetic measurements. Nucleation effect induced by the additive Fe enhances the coercivity ($H_c$) up to 27 kOe which is much larger than 4.5 kOe obtained for pure $SmCo_5$.

## 1. Introduction:

SmCo$_5$ is popular among all the rare-earth permanent magnets because of its usability in high temperature applications [1, 2]. Also hexagonal CaCu$_5$-type structure ensures high room temperature magnetic anisotropy [2, 3]. Density functional calculations for RCo$_{5-x}$Fe$_x$ [R = Rare-earth elements, Y, Sm] [4, 5] show increase in magnetic anisotropy as well as magnetization upon small fraction of Fe addition (x ≤ 0.35). However, very limited experimental work [6] has been reported so far to support these calculations. Increase in magnetic anisotropy is necessary in order to get high coercivity required for conventional hard magnetic applications. Replacement of cobalt by iron helps to improve the magnetization due to higher saturation magnetization of Fe and also decreases the cost of bulk magnets due to much lower cost of iron compared to cobalt. However, concentration of Fe in SmCo$_{5-x}$Fe$_x$ should be varied in a short interval to obtain a detailed insight of the physical process by studying the phase stability as well as change in magnetic properties of these alloys. In the present study, we have fabricated SmCo$_{5-x}$Fe$_x$ (0 ≤ x ≤ 2) alloys and studied their structure and magnetic properties. Noticeable improvement of magnetic properties is observed in SmCo$_{5-x}$Fe$_x$ due to the domain wall nucleation process induced by the presence of Fe.

## 2. Experimental Details:

Stoichiometrically weighed high purity Sm, Fe and Co elements are melted in argon atmosphere using arc-melting furnace. The melting is done repeatedly to get homogeneously mixed ingots. The weight loss after melting is less than 0.5% in each composition of SmCo$_{5-x}$Fe$_x$ (0 ≤ x ≤ 2). The arc melted ingots are taken in quartz crucible with an orifice of 0.5 mm and melted again using induction furnace in argon atmosphere. The melt of the ingot is shoot at high pressure through the orifice on a copper wheel kept at room temperature and rotating at a speed of 22 m/s which forms ribbons of SmCo$_{5-x}$Fe$_x$ and are collected in a chamber kept at room temperature. The melt spun ribbons are mechanically milled to powder for X-Ray diffraction (XRD, Rigaku D/Max-B x-ray diffractometer) studies. Magnetizations (*M*) vs. applied magnetic field (*H*) at 300K for the ribbons are obtained from the SQUID (Quantum Design Magnetic Properties Measurement System (MPMS) superconducting quantum interference device) measurements. To study the nucleation effect, non-magnetized samples are used to get the virgin magnetization curve. Compositions of the samples are investigated using energy dispersive x-ray analysis (EDX, JEOL JSM 840A scanning electron microscope).

## 3. Results and Discussions:

XRD measurements for the powdered $SmCo_{5-x}Fe_x$ are shown in fig. 1a and the peaks are in good agreement with the prototype hexagonal $CaCu_5$-type structure [2, 7] for x < 2. In the expanded XRD for x = 0, 0.5 and 0.75 shown in fig. 1b, the shift in the peak positions upon Fe addition (especially peaks (110) and (101)) indicated by the dotted lines confirms the lattice expansion due to replacement of Co by radially larger Fe atoms.

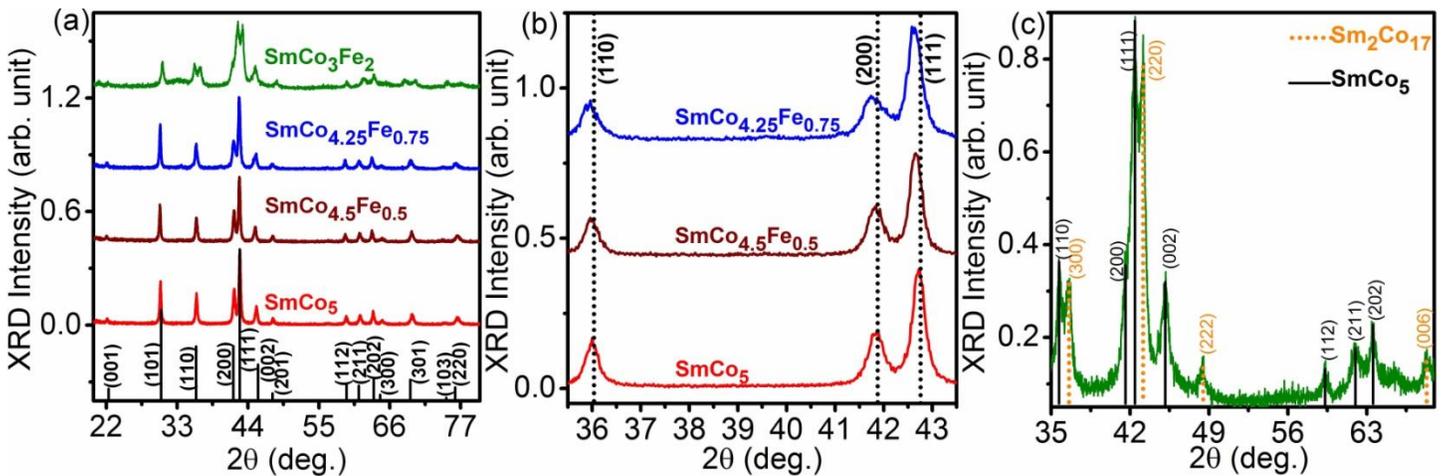

**Fig.1. a)** XRD of $SmCo_{5-x}Fe_x$ for x = 0, 0.5, 0.75 and 2 compared with prototype $CaCu_5$-structure (vertical lines). **b)** Expanded view of the XRD showing left shift of the peaks with respect to $SmCo_5$ (vertical dotted lines) due to Fe addition. **c)** XRD of $SmCo_3Fe_2$ showing mixture of $SmCo_5$ (vertical solid lines) and $Sm_2Co_{17}$ (vertical dotted lines) phases.

The grain size of the ribbons was determined using the well-known Scherrer formula [8, 9] and the calculation yields an average grain size of about 50 nm. Figure 1c shows mixture of $SmCo_5$ and $Sm_2Co_{17}$ phases [10] and the corresponding XRD peaks are denoted by solid and dotted lines respectively. The compositions of the samples were verified by EDX measurements.

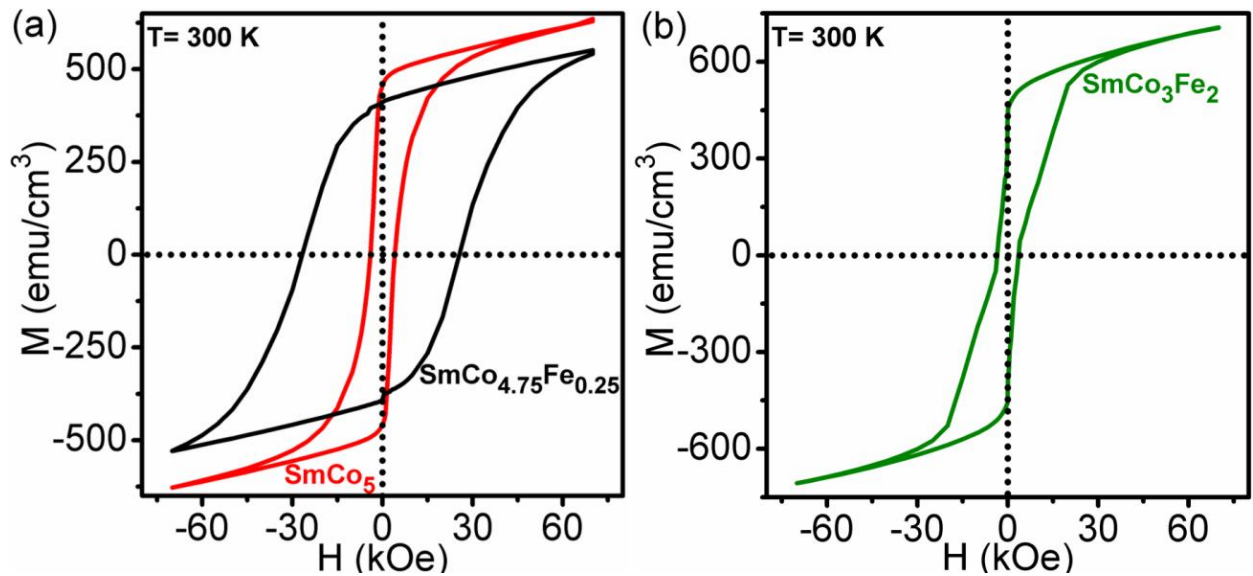

**Fig. 2. a)** Magnetization (*M*) vs. applied magnetic field (*H*) plots for $SmCo_5$ and $SmCo_{4.75}Fe_{0.25}$ showing improvement of coercivity with Fe addition. **b)** M-H hysteresis loop of $SmCo_3Fe_2$ exhibiting hard-soft two phase characteristic.

Figure 2a displays the hysteresis loops of the $SmCo_{5-x}Fe_x$ for x=0 and 0.25. $H_c$ of about 27 kOe is obtained for x = 0.25 which is an appreciable increase over 4.5 kOe corresponding to $SmCo_5$ (x=0) phase. As shown in fig. 2a, the hysteresis loops (at 300K) are not saturating even at 70 kOe applied magnetic field (*H*) which predicts a high magnetic anisotropy associated with the alloys. For x = 2, the hysteresis loop (fig. 2b) indicates mixture of phases which is in accordance with the XRD measurements.

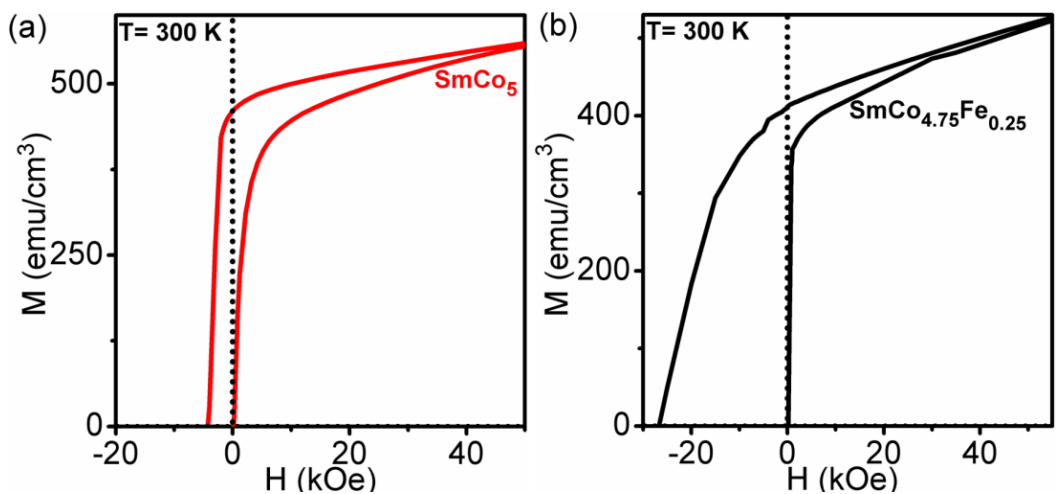

**Fig. 3.** Virgin *M* vs. *H* graph for **a)** $SmCo_5$ and **b)** $SmCo_{4.75}Fe_{0.25}$. The slope of the virgin plot region for b) is higher than a) showing domain wall nucleation effect induced by Fe addition.

Magnetization curve for the non-magnetized sample is referred as virgin hysteresis loop and is useful to determine the nature of domain wall motion due to either nucleation effect or pinning [11, 12]. In fig. 3 (a and

b) the virgin hysteresis loops of $SmCo_{5-x}Fe_x$ are shown for x=0 and 0.25 respectively. The slope of the virgin curve starting from zero applied magnetic field is higher for x = 0.25 (fig. 3b) in comparison with pure $SmCo_5$ (fig. 3a) which indicates the nucleation effect [11-13] upon Fe addition. Note that, pinning effect usually lowers the slope of the virgin curve due to the stiffness of the domain walls by the pinning centers which leads to low initial susceptibility [11, 14]. In case of nucleation, initial susceptibility is higher due to the free domain wall motion at initial (virgin) stage. Due to the domain wall nucleation, the magnetization reversal becomes hard as it is opposed by nucleated domains with magnetization opposite to the applied magnetic field [11, 14].

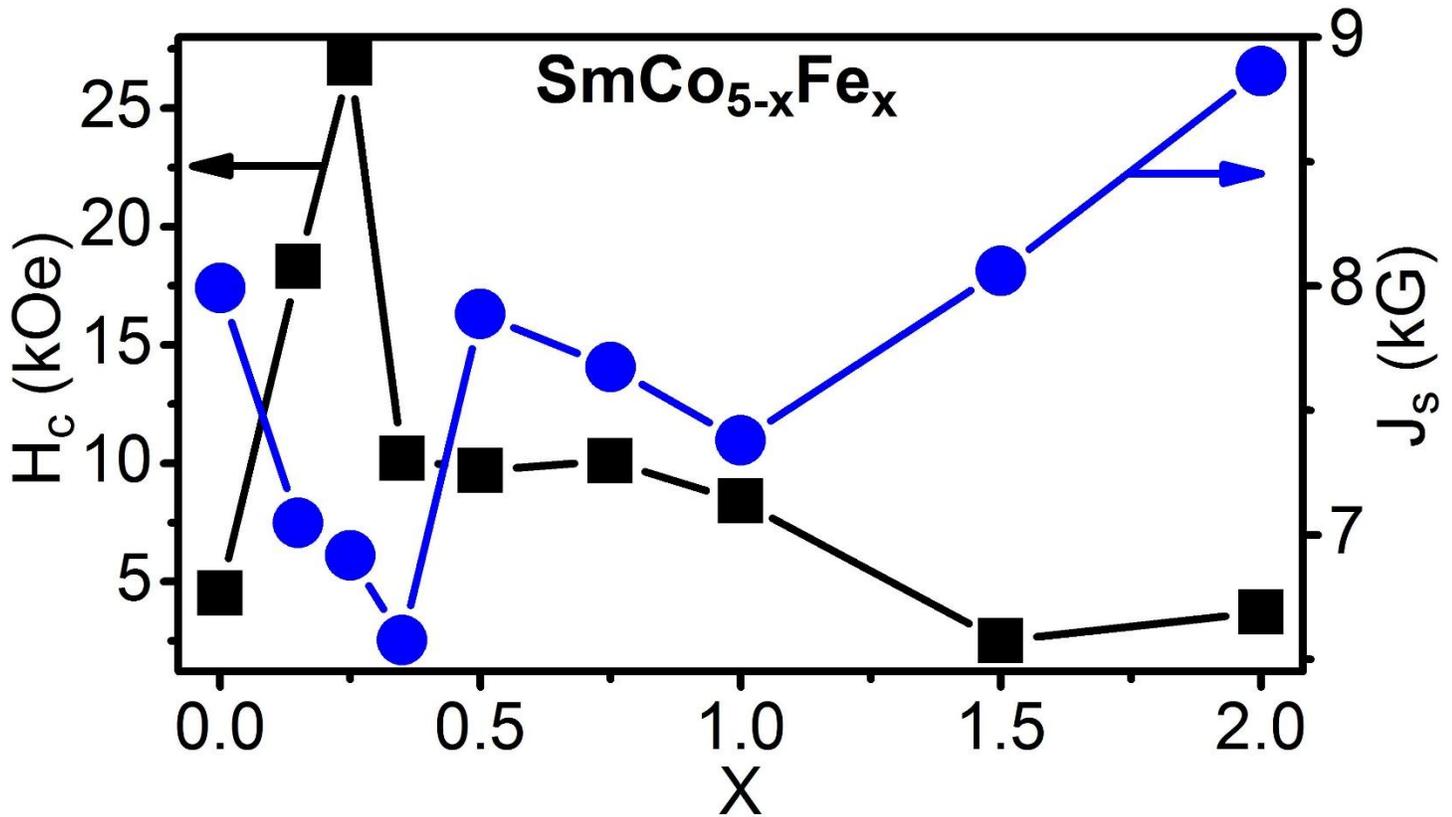

**Fig. 4.** Magnetic properties of $SmCo_{5-x}Fe_x$ for a range of x between 0 and 2.

Figure 4 summarizes the magnetic properties of $SmCo_{5-x}Fe_x$ for $0 \leq x \leq 2$. It is seen that the coercivity decreases after x = 0.25 possibly due to the weak nucleation effect at higher Fe concentration which is also observed in the virgin hysteresis loops for x > 0.25 (not shown here). Saturation polarization $J_s$ ($J_s = 4\pi M_s$, $M_s$ = saturation magnetization) increases for $x \geq 1$ due to the increase in Fe concentration.

The present results indicate, in addition to intrinsic effect of Fe on magnetocrystalline anisotropy, nanostructure and phase stability; the coercivity mechanism of $SmCo_{5-x}Fe_x$ is also controlled by induced nucleation effect. The reversal of magnetic domain requires large applied magnetic field and can be exploited to improve the hard-magnetic properties of the $SmCo_{5-x}Fe_x$ alloys suitable for applications.

## 4. Conclusion:

Changes in structural and magnetic properties of $SmCo_5$ magnets upon Fe addition has been studied with small increment of Fe concentration. The results show an appreciable range ($0 \leq x < 2$) of Fe concentration in $SmCo_{5-x}Fe_x$ which have hexagonal $SmCo_5$ – phase. A phase mixture occurs for $x \geq 2$ which is verified by XRD and magnetic measurements. The study of phase stability helps to determine the control of the magnetic properties related to the structural changes. Domain wall nucleation effect and increase in magnetic anisotropy are observed with increase in Fe concentration which leads to a noticeable improvement in magnetic properties with a maximum coercivity of about 27 kOe which is appreciable for magnetically unaligned $SmCo_5$–based alloys. This study is useful to understand the coercivity mechanism of the $SmCo_{5-x}Fe_x$ nanostructured alloys governed by domain wall motion and helps to tune the magnetic properties in order to obtain the optimum in the context of magnetic applications.

**Acknowledgements**

This work is supported by the US Department of Energy (Grant No. DE-FG02-04ER46152, D.J.S.), and Nebraska Center for Materials and Nanoscience (Central Facilities). Thanks are due to J.E. Shield, V.R. Shah, for instrument facilities and helpful discussions.